\documentclass[
]{iopart}
\usepackage[tight]{subfigure}
\usepackage{iopams}
\usepackage{epsfig}
\usepackage{color}
\usepackage{times}
\usepackage{graphics}
\newcommand{\sgpe}{stochastic Gross-Pitaevskii equation}
\newcommand{\GPE}{Gross-Pitaevskii equation}

\newcommand{\EQ}[1]{\begin{eqnarray}#1\end{eqnarray}}
\newcommand{\Label}[1]{\label{#1}}

\newcommand{\PP}{{\cal P}}
\newcommand{\QQ}{{\cal Q}}
\def\LABELS{
\renewcommand{\Label}[1]{\label{##1}
{\hbox to 0cm{\textcolor{blue}{\hss\em ##1\quad}}}}}
\begin{document}

\title[Properties of the SGPE]{Properties of the stochastic Gross-Pitaevskii equation: Projected Ehrenfest relations and the optimal plane wave basis}
\author{A. S. Bradley\dag, P. B. Blakie\ddag, and C. W. Gardiner\dag}
\address{\dag\ School of Chemical and Physical Sciences, Victoria
University of Wellington, New Zealand.}
\address{\ddag\ Physics Department, University of Otago, Dunedin, New Zealand}
\ead{ashton.bradley@gmail.com}
\date{\today}
\begin{abstract}
We investigate the properties of the \sgpe describing a 
condensate interacting with a stationary thermal cloud derived by Gardiner \etal
\cite{SGPE,SGPEII}. 
We find the appropriate Ehrenfest relations for the SGPE, 
including the effect of growth noise and projector terms arising from 
the energy cutoff. This is carried out in the high temperature regime appropriate for the SGPE of \cite{SGPEII}, which simplifies the action of
the projectors. 
\par
The validity condition for neglecting the projector terms in the Ehrenfest relations is found to be more stringent than
the usual condition of validity of the truncated Wigner method or classical field method \cite{Sinatra2002} -- which is that all 
modes are {\em highly occupied}.
In addition it is required that the overlap of the nonlinear term with the lowest energy eigenstate of the non-condensate band is small. We show how to use the Ehrenfest relations along with the corrections generated by the projector to monitor dynamical artifacts arising from the cutoff.
\par
We also investigate the effect 
of using different bases to describe a harmonically trapped BEC at finite
temperature by comparing the condensate fraction found using the plane wave and single particle bases.
We show that the equilibrium properties are strongly dependent 
on the choice of basis. There is thus an optimal choice of plane wave basis for a given cut-off energy and we show that this basis gives the best reproduction of the single
particle spectrum, the condensate fraction and the position and momentum densities. 
\end{abstract}


\section{Introduction}
Since the experimental achievement of Bose-Einstein condensation
\cite{Anderson1995}, the
theoretical tool of choice for describing condensates is the \GPE \;(GPE) 
\cite{Dalfovo1999}, which has been found to describe a remarkably wide range of ultra-low temperature
BEC physics. The extension of the GPE to finite temperatures in recent years 
\cite{Marshall1999,Davis2001a,Davis2001b,Davis2002,Davis2003} approaches 
the problem by treating the highly
occupied modes of the condensate as a multimode `classical field'. This has developed 
alongside various stochastic generalisations of the GPE based on the phase space
methods of quantum optics \cite{Steel1998,Stoof1999,Stoof2001}. The recent 
formulation of Gardiner and coworkers \cite{SGPE,SGPEII} derives
a \sgpe (SGPE) for the condensate band of a partially condensed Bose gas in
contact with the non condensed thermal cloud. The treatment uses the
truncated Wigner approximation (TWA) \cite{QO,Graham1973} which has been used with a certain
degree of success to describe highly occupied optical fields for many years \cite{TWAlimits}; 
however, even in quantum optics, where the occupation numbers of the modes of interest
are enormous, this approach has its limitations.
\par
The validity of the TWA for partially Bose condensed gases has recently been investigated
in some detail by Sinatra {\em et al} \cite{Sinatra2002}. The authors 
model a trapped finite temperature Bose-Einstein condensate using the TWA
and a plane wave basis; consequently, the method uses a momentum 
cut-off in the numerical representation of the quantum field. In the high temperature regime it was shown
that the condition of validity for the TWA is that all modes must be highly occupied.
The approach of Gardiner and Davis \cite{SGPEII} is to use a GPE that is 
projected into a basis that generates a strict energy cut-off for the harmonic trap, for which all 
modes are highly occupied in the low energy region. The weakly occupied modes above the cut-off 
are then traced out, generating noise terms in the stochastic Gross-Pitaevskii 
equation (SGPE). An important computational question then arises: What are the Ehrenfest equations for 
the SGPE, and under what conditions can we recover physical behaviour that resembles the familiar motion of the GPE? This will be addressed in the first part of this paper, where it is
shown that when the spatial variation of the thermal cloud is neglected the GPE Ehrenfest equations can be extended to the SGPE. We derive exact expressions for the boundary terms in the Ehrenfest equations generated by the projector which we use to devise a method of assessing the influence of the projector on the dynamics -- a method which we expect to be applicable under the conditions of the TWA.
\par
The second part of the paper considers the validity of using a plane wave basis to 
model a harmonically trapped condensate. Until very recently 
numerical methods based on plane wave representations have been favoured largely
because of the speed advantage gained by using pseudo-spectral
Fourier transform methods \cite{pseudoSpec}, and the conceptual ease of use that is inherent in the
plane wave basis. However, there are a number of approximation 
issues that arise when using a plane wave basis to describe a trapped condensate at {\em finite
temperature} which have not previously been investigated. We explore the link between the 
TWA or classical field approximation and the 
basis of representation in detail. To determine how best the
plane wave basis may be used for a trapped BEC we compare the plane wave
basis with an efficient spectral method based on numerical quadrature developed by Dion and
Cances \cite{Dion2003}, and applied to the finite temperature GPE by Blakie and Davis \cite{Blakie2004A}.
We show how to construct the optimal plane wave basis for a given cut-off, and verify that this
basis gives the best agreement with the spectral basis when computing the condensate fraction for
a partially condensed Bose gas. Variation of the basis generates significantly different
condensate fractions, and increases the region of phase space that is not modeled accurately by
the basis. In any representation this region of phase space must be 
significantly occupied due to the TWA validity condition, but it is minimized for the optimal
plane wave basis. 
\par
Although the spectral method is more difficult to implement and often more computationally expensive than plane wave 
approaches \cite{SlowSpec}, there are
clear advantages to using it to describe a trapped finite temperature condensate.
The energy cut-off is well defined, and the high occupation condition can be imposed
symmetrically in the phase space. There is another important feature that is basis dependent: the
form of the Ehrenfest equation for the condensate band energy depends on the basis in
which it is evaluated. Using the correct basis numerically thus becomes vital if one wants to
obtain the correct dynamics for the condensate band of an open system using the TWA.

\section{Background}
It is well known that a solution of the GPE will satisfy the same Ehrenfest relations 
as solutions of the Schr\"odinger equation \cite{BMCDThesis}. We briefly reiterate these here to
establish notation.

\subsection{Ehrenfest relations for the Gross-Pitaevskii equation}
The GPE is the equation of motion for a field evolving according to the Hamiltonian functional
\EQ{\fl
\Label{Hdef}H[\psi,\psi^*]=\int
d^3\mathbf{x}\;\psi^*(\mathbf{x},t)\left(-\frac{\hbar^2\nabla^2}{2m}+V(\mathbf{x},t)+\frac{u}{2}|\psi(\mathbf{x},t)|^2\right)\psi(\mathbf{x},t),
}
which is obtained by differentiating:
\EQ{\Label{GPEdef}
i\hbar\frac{\partial \psi}{\partial t}=\frac{\delta H}{\delta \psi^*}=L_{GP}\psi,
}
where the Gross-Pitaevskii operator is
\EQ{\Label{Lop}L_{GP}\psi({\bf
x})\equiv\left(-\frac{\hbar^2\nabla^2}{2m}+V(\mathbf{x},t)+u|\psi({\bf x})|^2\right)\psi({\bf x}).
}
In this paper $V({\bf x},t)$ describes a general time dependent trapping potential, and $u\equiv
4\pi\hbar^2a/m$ is the S-wave limit interaction strength \cite{Dalfovo1999}.
The quantities of interest are the energy density
\EQ{
U=-\frac{\hbar^2\nabla^2}{2m}+V(\mathbf{x},t)+\frac{u}{2}|\psi(\mathbf{x},t)|^2,
}
position, momentum and angular momentum (${\bf x}, {\bf p}=-i\hbar\nabla$ and ${\bf L}=-i\hbar\mathbf{x}\times\nabla$ respectively).
The Ehrenfest equations are
\EQ{
\Label{x}\frac{d\langle \mathbf{x}\rangle}{dt}=\frac{\langle \mathbf{p}\rangle}{m}\\
\Label{p}\frac{d\langle \mathbf{p}\rangle}{dt}=-\langle\nabla V\rangle\\
\Label{L}\frac{d\langle \mathbf{L}\rangle}{dt}=-\frac{i}{\hbar}\langle\mathbf{L}V\rangle\\
\Label{U}\frac{d\langle U\rangle}{dt}=\Big\langle\frac{\partial V}{\partial t}\Big\rangle,
}
where $\langle A\rangle\equiv\int d{\bf x}\; \psi^*A\psi$, and
the energy is simply $\langle U\rangle=H[\psi,\psi^*]$. We are working with the many particle wavefunction, so for simplicity of notation the condensate band occupation number will be written as
$\langle N \rangle\equiv\int d{\bf x}\; \psi^*\psi$.
\subsection{Properties of projectors}\Label{sec:Pproperties}
Projectors that carry out the separation into upper and lower energy bands are defined 
by first separating the potential 
\EQ{V(\mathbf{x},t)\equiv V_0(\mathbf{x})+\delta V(\mathbf{x},t),}
where the time invariant part is used to define the single particle Hamiltonian
\begin{equation}
H_0\equiv\frac{-\hbar^2\nabla^2}{2m}+V_0(\mathbf{x}),
\end{equation}
and $\delta V(\mathbf{x},t)$ is arbitrary.
The representation basis is provided by the eigenstates 
\EQ{\Label{H0def}
H_0\;\phi_n(\mathbf{x})=\epsilon_n\;\phi_n(\mathbf{x}).}
We introduce orthogonal projection operators which are defined with respect to the single particle basis by their action on an arbitrary wavefunction $\psi(\mathbf{x})$
\EQ{
\Label{Pdef}\PP\psi\equiv\sum_{n\leq N_c} \phi_n(\mathbf{x})\int
d^3\mathbf{y}\;\phi^*_n(\mathbf{y})\psi(\mathbf{y})
}
\EQ{
\Label{Qdef}\QQ\psi\equiv\sum_{n> N_c} \phi_n(\mathbf{x})\int
d^3\mathbf{y}\;\phi^*_n(\mathbf{y})\psi(\mathbf{y}),
}
and satisfy $\PP+\QQ=1$
The index
of summation $n$ represents all the eigenvalues required to specify the complete
set of single particle modes, and $N_c$ defines the cut-off energy. For convenience the cut-off is chosen so that $N_c$ as an integer; hereafter we will use the notation
\EQ{\Label{Psum}
\sum_{n\leq N_c}\equiv\sum_n^{-}
}
for projected sums over the condensate band.
\par
We now restrict our attention to the condensate band itself so that $\psi\equiv\PP\psi$ for the wavefunctions of interest.
The projected GPE (PGPE) corresponding to \eref{GPEdef} can be written as
\EQ{
\Label{Qdpsidt}\frac{\partial \psi}{\partial t}=-\frac{i}{\hbar}(1-\QQ) L_{GP}\psi.}
By writing the PGPE in terms of the projector orthogonal to $\PP$ we can use the properties
\EQ{
\Label{Pstar}(\QQ\psi)^*&=&\QQ^*\psi^*\\
\Label{Padjoint}\int d^3\mathbf{x}\;\phi^*\QQ\psi&=&\int d^3\mathbf{x}\;(\QQ\phi)^*\psi\\
\Label{PHo}\QQ\PP\psi&=&\QQ\psi=\QQ H_0\psi=0
}
to extract the Ehrenfest relations for the condensate band wavefunction along
with modifications that arise from the $\QQ$ projector.
We make use of these relations in \sref{sec:Ehrenfest}. We have also used the following notation for the complex 
conjugate projector
\EQ{
\Label{Pstardef}\QQ^*\psi\equiv\sum_{n> N_c} \phi^*_n(\mathbf{x})\int
d^3\mathbf{y}\;\phi_n(\mathbf{y})\psi(\mathbf{y}).
}
The delta function restricted to the condensate band 
\EQ{\Label{deltadef}
\delta_C(\mathbf{x},\mathbf{y})\equiv\sum_n^-\phi_n(\mathbf{x})\phi_n^*(\mathbf{y})
}
has the projection property
\EQ{
\int d^3\mathbf{y}\;\delta_C(\mathbf{x},\mathbf{y})f(\mathbf{y})=\PP f(\mathbf{x})
}
for any function $f$. $\delta_C(\mathbf{x},\mathbf{y})$ behaves like a true delta distribution for functions restricted to the condensate band:
\EQ{
\int d^3\mathbf{y}\; \delta_C(\mathbf{x},\mathbf{y})\PP\psi(\mathbf{y})\equiv\PP \psi(\mathbf{x}),
}
which is equivalent to $\PP\PP=\PP$.
Note that a straightforward application of \eref{Padjoint} shows that the PGPE is energy and number conserving.
\section{The stochastic Gross-Pitaevskii equation}
The SGPE formalism \cite{SGPEII} separates the partially condensed system
into a low energy subspace of modes, or {\em condensate band},
and its orthogonal complement , the union of which furnishes a complete basis. The 
{\em non-condensate band} is assumed thermalized, so that it may be described by Gaussian statistics and traced out.  The
non-condensate band thus plays the role of a thermal reservoir and acts as a damping mechanism for the condensate band, while the condensate band contains the condensate and its excitations, along with a low energy thermal component.
\par
While this description is somewhat more complicated 
that the PGPE
\cite{Davis2001a,Davis2001b,Davis2002,Davis2003} when treated in full, if we neglect the
scattering term (which does affect the condensate band occupation) and take the limit 
of a broad thermal cloud, it can be reduced to a relatively simple equation of motion 
for the condensate band which is closely related to the PGPE. The complete derivation may be found in reference \cite{SGPEII}, but here we will briefly sketch the derivation, with a few minor changes 
of notation to make a transparent connection with the results to follow. 
\par
To obtain the 
SGPE we proceed from the second quantised Hamiltonian 
for the system in the S-wave scattering limit
\EQ{
\fl H=\int
d^3\mathbf{x}\;\Psi^\dag(\mathbf{x})\left(-\frac{\hbar^2\nabla^2}{2m}+V(\mathbf{x},t)\right)\Psi(\mathbf{x})+\frac{u}{2}\int
d^3\mathbf{x}\;\Psi^\dag(\mathbf{x})\Psi^\dag(\mathbf{x})\Psi(\mathbf{x})\Psi(\mathbf{x}).
}
The field operator is split at the cut-off energy into $\Psi(\mathbf{x})=\phi(\mathbf{x})+\psi_{\rm NC}(\mathbf{x})$, where the non-condensate
field $\psi_{\rm NC}(\mathbf{x})$ describes the high energy thermal modes. The commutator of the condensate band field operator is 
\EQ{
[\phi(\mathbf{x}),\phi^\dag(\mathbf{y})]=\delta_C(\mathbf{x},\mathbf{y}).
}
The thermal statistics of the non condensate field allow averages over many non-condensed 
field operators to be factorised and reduced to products of single particle Wigner 
functions $F(\mathbf{u},\mathbf{v})$. In this work we neglect the phase damping processes 
which lead to the `scattering' terms in the master equation of \cite{SGPEII}. The 
growth/loss master equation for the reduced density matrix of the condensate band 
$\rho_{C}={\rm Tr_{NC}}(\rho)$ can be written in terms of the amplitudes\footnote[1]{This corrects 
an extra minus sign in the defining equation (56) of \cite{SGPEII}.} 
\EQ{\fl
\Label{Gplus}G^{\scriptscriptstyle(+)}(\mathbf{u},\mathbf{v},\epsilon)=\frac{u^2}{(2\pi)^8\hbar^2}\int d^3\mathbf{K}_1\int d^3\mathbf{K}_2\int d^3\mathbf{K}_3\;F(\mathbf{u},\mathbf{K}_1)F(\mathbf{u},\mathbf{K}_2)[1+F(\mathbf{u},\mathbf{K}_3)]\nonumber\\
\lo\times\delta(\omega_1+\omega_2-\omega_3-\epsilon/\hbar)e^{-i(\mathbf{K}_1+\mathbf{K}_2-\mathbf{K}_3)\cdot\mathbf{v}}
}
and
\EQ{
\Label{Gminus}G^{\scriptscriptstyle(-)}(\mathbf{u},\mathbf{v},\epsilon)=e^{(\epsilon-\mu)/k_BT}G^{\scriptscriptstyle(+)}(\mathbf{u},\mathbf{v},\epsilon)
}
in the form\footnote[2]{This corrects an error in equation (59) of \cite{SGPEII} wherein $L_C$ appeared in place
of $-L_C$ in the second and third lines of the master equation equivalent to \eref{rhogrowth}}
\EQ{
\fl\Label{rhogrowth}\dot{\rho}_C|_{\rm growth}=\;\;\;\;\;\;\;\;\int d^3\mathbf{u}\int d^3\mathbf{v}\left[\left\{G^{\scriptscriptstyle(-)}(\mathbf{u},\mathbf{v},L_{ C})\phi(\mathbf{u}-\mathbf{v}/2)\right\}\rho_{C},\phi^\dag(\mathbf{u}+\mathbf{v}/2)\right]\nonumber\\
\lo-\int d^3\mathbf{u}\int d^3\mathbf{v}\left[\rho_{ C}\left\{G^{\scriptscriptstyle(-)}(\mathbf{u},\mathbf{v},-L_{ C})\phi^\dag(\mathbf{u}-\mathbf{v}/2)\right\},\phi(\mathbf{u}+\mathbf{v}/2)\right]\nonumber\\
\lo+\int d^3\mathbf{u}\int d^3\mathbf{v}\left[\left\{G^{\scriptscriptstyle(+)}(\mathbf{u},\mathbf{v},-L_{ C})\phi^\dag(\mathbf{u}-\mathbf{v}/2)\right\}\rho_{C},\phi(\mathbf{u}+\mathbf{v}/2)\right]\nonumber\\
\lo-\int d^3\mathbf{u}\int d^3\mathbf{v}\left[\rho_{ C}\left\{G^{\scriptscriptstyle(+)}(\mathbf{u},\mathbf{v},L_{ C})\phi(\mathbf{u}-\mathbf{v}/2)\right\},\phi^\dag(\mathbf{u}+\mathbf{v}/2)\right].
}
The condensate band operator is given in terms of the condensate band Hamiltonian
\EQ{\fl
H_{C}=\int d^3\;\mathbf{x}\;\phi^\dag(\mathbf{x})\left(-\frac{\hbar^2\nabla^2}{2m}+V(\mathbf{x},t)\right)\phi(\mathbf{x})
+\frac{u}{2}\int d^3\mathbf{x}\;\phi^\dag(\mathbf{x})\phi^\dag(\mathbf{x})\phi(\mathbf{x})\phi(\mathbf{x})
}
as
\EQ{\Label{Lcfielddef}
L_{C}\phi(\mathbf{x})&\equiv&[\phi(\mathbf{x}),H_{C}].
}
In principle a mean field or forward scattering term could also be included in $H_{C}$, and would alter the effective potential. In what follows we account for this possibility by including a general time dependent perturbing potential, into which such a term can be absorbed. 

The master equation for the growth is found by 
\begin{itemize}
\item[i)] Expanding the exponential in the forward-backward relation \eref{Gminus} to first order in $(\epsilon-\mu)/k_BT$ 
\item[ii)] Neglecting the condensate band energy during collisions with thermal atoms 
by approximating $G^{\scriptscriptstyle(+)}(\mathbf{x},\mathbf{v},\epsilon)\approx G^{\scriptscriptstyle(+)}(\mathbf{x},\mathbf{v},0)$
\item[iii)] Neglecting the condensate band momentum by making the approximation $\phi(\mathbf{u}\pm \mathbf{v}/2)\approx \phi(\mathbf{u})$
\end{itemize}
(i) is valid as long as the condensate band chemical potential is not significantly 
different from the non condensate chemical
potential. This can be satisfied for a wide range of temperatures, and should be true 
for most physical situations of interest -- the resulting SGPE is a valid description 
whenever the energy fluctuations between the two bands are small relative to $k_BT$. A more 
detailed model of non condensate 
dynamics is required for this to be consistent in general since the dynamics of the non-condensate band should also be treated, but for many situations of interest
the non condensate band can be described by a stationary distribution. Indeed, conditions 
(ii) and (iii) neglect energy and momentum exchange between the two bands, 
so the resulting 
equation is self consistent as long as the thermal cloud can be expected to remain 
stationary throughout the motion. Approximation (iii) can be made because 
$G^{(+)}(\mathbf{u},\mathbf{v},0)$ is sharply peaked about $\mathbf{v}=0$. We further treat the 
growth amplitude $G^{(+)}(\mathbf{u},\mathbf{v},0)$ as spatially constant. This 
is the main simplifying approximation of the present work, and corresponds to the high temperature
regime where the thermal cloud density is approximately constant over the condensate band.
The growth parameter becomes
\EQ{
\gamma=\frac{\bar{G}(0)}{k_BT}\equiv\frac{1}{k_BT}\int d^3\mathbf{v}\;G^{\scriptscriptstyle(+)}(0,\mathbf{v},0).
}
Relaxing this assumption is possible in principle, but leads to Ehrenfest relations that have a complicated 
dependence on the shape of the thermal cloud and the cut-off energy of the projector. The form we use simplifies the projection as much as possible. 
\par 
The linearization required to obtain the SGPE from \eref{rhogrowth} is an expansion in the 
operator $\gamma(L_C-\mu)/k_BT$, which requires 
this to be small compared to the usual Gross-Pitaevskii evolution. The prefactor is 
usually of the order $\hbar\gamma \sim 10^{-3}$ so 
this is easily satisfied in practice.
In the high temperature regime $\gamma$ takes the simple form
\EQ{\Label{gammadef}
\gamma=\frac{16k_BTa^3}{\hbar u},
}
which we will use in this work.
\par
The SGPE derivation then follows the familiar path of quantum optics (see \cite{SGPEII}): 
The master equation for the reduced density matrix of the condensate band is mapped onto an equation 
of motion for the multimode Wigner distribution. Derivatives higher than second order in the fields 
are neglected in order to derive a genuine Fokker-Planck equation with positive definite diffusion 
matrix. This is mapped to a stochastic differential equation for the condensate band field 
$\alpha(\mathbf{x},t)$ \cite{QN}. The distinction between \cite{SGPEII} and earlier work is a rigorous formulation 
in terms of projection operators that generate a consistent energy cut-off for the condensate band.
\par 
We can now write the the SGPE as the It\^o stochastic differential equation
\EQ{\fl
\Label{sgpedef}d\alpha(\mathbf{x},t)=-\frac{i}{\hbar}\PP L_{GP}\alpha(\mathbf{x},t)dt+\gamma\PP(\mu-L_{GP})\alpha(\mathbf{x},t)dt
+dW_G(\mathbf{x},t),
}
where the noise is a vector Wiener process which satisfies
\EQ{\Label{n1}
dW_G(\mathbf{x},t)dW_G(\mathbf{x}^\prime,t)&=&dW_G^*(\mathbf{x},t)dW_G^*(\mathbf{x}^\prime,t)=0\\
\Label{n2}
dW_G^*(\mathbf{x},t)dW_G(\mathbf{x}^\prime,t)&=&2\gamma k_BT\delta_C(\mathbf{x},\mathbf{x}^\prime)dt,
}

An important consequence of neglecting the scattering in \cite{SGPEII} is that the growth 
noise is purely additive, so that \eref{sgpedef} is the same in It\^o or Stratonovich form \cite{SM}. 
\subsection{The PGPE and Fudge equations}
Equation \eref{sgpedef} can be used to recover two useful equations which have already 
been used to successfully describe a number of interesting finite temperature effects in BEC 
including equilibrium properties and vortex lattice nucleation
\cite{Davis2001a,Davis2001b,Davis2002,Davis2003,Penckwitt2002,Tsubota2002}.  

Putting $\gamma=0$ recovers the PGPE \eref{Qdpsidt} 
\EQ{
\Label{PGPE}i\hbar\frac{\partial \alpha(\mathbf{x},t)}{\partial t}=\PP\left\{\left(-\frac{\hbar^2\nabla^2}{2m}+V(\mathbf{x},t)+u|\alpha(\mathbf{x},t)|^2\right)\alpha(\mathbf{x},t)\right\}.
}
The projector is number conserving and acts to keep the wavefunction in a restricted 
energy subspace of the trap. 
The careful addition of this single feature to the GPE expands its 
scope to finite temperatures, where the field $\alpha(\mathbf{x},t)$ 
provides a non-perturbative description of both the condensate and a range of excitations up to the cut-off energy of the projector \cite{MJDThesis}.
\par
A rigorous form of the phenomenological finite temperature GPE 
-- or Fudge equation~\cite{SGPE,Penckwitt2002}\ -- is found from \eref{sgpedef} by dropping the noise and settting the projector to the identity, to give
\EQ{\fl\Label{GAF}
i\hbar\frac{\partial \alpha(\mathbf{x},t)}{\partial t}=(1-i\hbar\gamma)\left(-\frac{\hbar^2\nabla^2}{2m}+V(\mathbf{x},t)+u|\alpha(\mathbf{x},t)|^2\right)\alpha(\mathbf{x},t)+i\hbar\gamma\mu\alpha(\mathbf{x},t).
}
This is a semiclassical equation for a condensate in contact with a thermal cloud at 
chemical potential $\mu$. Although many approximations have been made to derive this equation they are all well defined, and it must not be
overlooked that this is therefore the logical equation to use when treating a damped condensate -- rather than the phenomenological approaches to
damping that have thus far been used \cite{Choi1998}. We also note that numerical integration of \eref{GAF} will lead to damping for {\em any} $\gamma$, rather than the small $\gamma$ limit required for the Fudge equation of \cite{SGPEII}.
\section{Continuity and Ehrenfest relations at finite temperature}\label{sec:Ehrenfest}
The projector can be dealt with by noting that, by putting $\PP=1-\QQ$ in
\eref{sgpedef}, the standard Ehrenfest and continuity results can be
recovered from the terms multiplied by the identity, and the properties
\eref{Pstar}\;--\;\eref{PHo} can be used to find the influence of the $\QQ$ projector.
\subsection{Continuity}
Averaging over the noise using It\^o rules leads to the continuity equation
\EQ{\fl
\frac{\partial n_C(\mathbf{x})}{\partial t}+\nabla\cdot\mathbf{j}_C(\mathbf{x})=&\frac{2}{\hbar}{\rm Im}\langle\QQ^*\left\{(\delta V(\mathbf{x},t)+u|\alpha(\mathbf{x})|^2)\alpha^*(\mathbf{x})\right\}\alpha(\mathbf{x})\rangle_{\scriptscriptstyle W}\nonumber\\
\fl&+\gamma2{\rm Re}\langle\QQ^*\left\{[\delta V(\mathbf{x},t)+u|\alpha(\mathbf{x})|^2]\alpha^*(\mathbf{x})\right\}\alpha(\mathbf{x})\rangle_{\scriptscriptstyle W}\nonumber\\
\fl\Label{cont1}&+\gamma2{\rm Re}\langle\mu |\alpha(\mathbf{x})|^2-\alpha^*(\mathbf{x})L_{GP}\alpha(\mathbf{x})\rangle_{\scriptscriptstyle W}
}
where we have used \eref{n2}, and $\langle \rangle_{\scriptscriptstyle W}$ denotes the Wigner average. The density and current take the usual form in terms of the wavefunction
$\alpha(\mathbf{x})$ and correspond to symmetrised operator averages for the condensate band 
\EQ{\fl
n_C(\mathbf{x})=\frac{1}{2}\langle\phi^\dag(\mathbf{x})\phi(\mathbf{x})+\phi(\mathbf{x})\phi^\dag(\mathbf{x})\rangle\\
\fl\mathbf{j}_C(\mathbf{x})=\frac{i\hbar}{4m}\langle\nabla\phi^\dag(\mathbf{x})\phi(\mathbf{x})-\phi^\dag(\mathbf{x})\nabla\phi(\mathbf{x})+\nabla\phi(\mathbf{x})\phi^\dag(\mathbf{x})-\phi(\mathbf{x})\nabla\phi^\dag(\mathbf{x})\rangle,
}
where the angle brackets around an operator expression denote the trace over the density matrix of the condensate band. 
When there is no damping ($\gamma=0$) the continuity equation for the resulting PGPE has 
an additional source term (the first term on the right hand side of equation (\ref{cont1})) which redistributes the field. However, for any function 
$f(\mathbf{x})$ and projected wavefunction $\alpha(\mathbf{x})$ 
\EQ{
\int d^3\mathbf{x}\;\QQ^*[f(\mathbf{x})]\alpha(\mathbf{x})=\int d^3\mathbf{x}\;f(\mathbf{x})\QQ\alpha(\mathbf{x})=0,
}
and consequently the source conserves atom number. The damping 
generates qualitatively different terms responsible for growth and fluctuations, as well as
another number conserving source term.
\subsection{Ehrenfest relations}
The fluctuation terms have a generic form that can be written as a trace over the projected 
operator; the energy has to be treated slightly differently because of the nonlinear term, and 
may not be cast in the same form as the other averages.
\par
We have overloaded the $\langle \rangle$ notation somewhat, since we wish to represent both spatial integration and stochastic averages to write down the Ehrenfest equations; in what follows we will adopt the convention 
\EQ{
\langle f\rangle_{\scriptscriptstyle W}\equiv \int d^3\mathbf{x}\;\langle\alpha^*(\mathbf{x}) f(\mathbf{x})\alpha(\mathbf{x})\rangle_{\scriptscriptstyle W},
}
that is, we will suppress the coordinates of spatial integration except where they are required to specify the operators which are being integrated. 
To carry out the averaging, we use It\^o rules to average over the noise, and the properties 
of the $\QQ$ projector. In the equation of motion for $\langle A\rangle$, where $A$ is one 
of the operators $\mathbf{x}, \mathbf{p}$ or $\mathbf{L}$, the It\^o correction comes about from 
the last term in
\EQ{
d\langle A\rangle=\int d^3\mathbf{x}\;\alpha^*(\mathbf{x})Ad\alpha(\mathbf{x})+d\alpha^*(\mathbf{x})A\alpha(\mathbf{x})+d\alpha^*(\mathbf{x})A d\alpha(\mathbf{x}),
}

\par
Carrying out the spatial and stochastic averaging leads to the Ehrenfest relations for 
the projected stochastic Gross-Pitaevskii equation:
\EQ{
\fl\Label{dNdtsgpe}\frac{d \langle N\rangle_{\scriptscriptstyle W}}{dt}&=&2\gamma\langle
\mu-L_{GP}\rangle_{\scriptscriptstyle W}+2\gamma k_BT\Tr(\PP),\\
\fl\Label{dxdtsgpe}\frac{d\langle\mathbf{x}\rangle_{\scriptscriptstyle W}}{dt}&=&\frac{\langle\mathbf{p}\rangle_{\scriptscriptstyle W}}{m}+2\gamma{\rm Re}\langle \mathbf{x}(\mu-L_{GP})\rangle_{\scriptscriptstyle W}+2\gamma k_BT\Tr(\PP\mathbf{x})+Q_{\mathbf{x}},\\
\fl\Label{dpdtsgpe}\frac{d\langle\mathbf{p}\rangle_{\scriptscriptstyle W}}{dt}&=&-\langle\nabla V\rangle_{\scriptscriptstyle W}+2\gamma{\rm Re}\langle \mathbf{p}(\mu-L_{GP})\rangle_{\scriptscriptstyle W}+2\gamma k_BT\Tr(\PP\mathbf{p})+Q_{\mathbf{p}},\\
\fl\Label{dLdtsgpe}\frac{d\langle
\mathbf{L}\rangle_{\scriptscriptstyle W}}{dt}&=&-\frac{i}{\hbar}\langle\mathbf{L}V\rangle_{\scriptscriptstyle W}
+\gamma2{\rm Re}\langle \mathbf{L}(\mu-L_{GP})\rangle_{\scriptscriptstyle W}+2\gamma k_BT\Tr(\PP\mathbf{L})+Q_{\mathbf{L}},\\
\fl\frac{d\langle U\rangle_{\scriptscriptstyle W}}{dt}&=&\left\langle\frac{\partial V}{\partial
t}\right\rangle_{\scriptscriptstyle W}+2\gamma\langle
\overleftarrow{L}_{GP}(\mu-L_{GP})\rangle_{\scriptscriptstyle W}+2\gamma k_BT(\Tr[\PP(H_0+\delta V)]+u\langle\delta_C\rangle),\nonumber\\
\fl\Label{dEdtsgpe}
}
where the left acting operator is
\EQ{
\alpha^*({\bf x})\overleftarrow{L}_{GP}\equiv L_{GP}\alpha^*({\bf x}),
}
and the boundary terms take the form
\EQ{\Label{QAdef}
Q_{A}&=&\frac{1}{\hbar}2{\rm Im}\langle F_A\rangle_{\scriptscriptstyle W}+\gamma2{\rm Re}\langle F_A\rangle_{\scriptscriptstyle W},
}
where
\EQ{\Label{Fadef}
F_A\alpha({\bf x})\equiv(\delta
V(\mathbf{x},t)+u|\alpha(\mathbf{x})|^2)\QQ [A\alpha({\bf x})].
}
\par
These equations serve as useful consistency conditions for numerical simulations of
the SGPE, and extend the Ehrenfest results to the projected formalism. The 
Ehrenfest behaviour familiar from the 
Gross-Pitaevskii equation is modified by growth terms and boundary corrections generated 
by the $\QQ$ projector. 
\par
It is worth noting that corresponding results for the Fudge equation \eref{GAF} are obtained by putting $T=0$ in equations \eref{dNdtsgpe}--\eref{dEdtsgpe},
while retaining $\gamma$ and omitting the noise average. Although, from \eref{gammadef} we see that $\gamma$ is in fact proportional to $T$, the equations have been 
cast so that $T$ only occurs explicitly in terms arising from the noise. 
It is clear from \eref{QAdef} that when $T=0$ the boundary term caused by
damping can still play a role in the Fudge equation dynamics. 
If we further put $\gamma=0$ the
results for the PGPE are recovered, which still have boundary corrections corresponding 
to the imaginary part of the source term in equation \eref{cont1}. 

\subsubsection{General remarks}\Label{sec:remarks}
\begin{itemize}
\item[i)]A typical trace term is, for example
\EQ{
{\rm Tr}(\PP\mathbf{x})={\rm Tr}\sum_n^-|\phi_n\rangle\langle \phi_n|\mathbf{x}=\sum_n^-\int d^3\mathbf{x}\;\phi_n^*(\mathbf{x})\mathbf{x}\phi_n(\mathbf{x})
}
The assumption of a homogeneous non-condensate thus leads to state independent driving from 
the projected operators; the exception is the energy equation \eref{dEdtsgpe}, where the 
last term
\EQ{
\langle \delta_C\rangle\equiv\int d^3\mathbf{x}\;\alpha^*(\mathbf{x})\delta_C(\mathbf{x},\mathbf{x})\alpha(\mathbf{x})
}
arises from the nonlinear interaction and cannot be cast as a trace. 

The term in \eref{dNdtsgpe}
\EQ{
\Tr(\PP)=\sum_{n}^-\langle \phi_n|\phi_n\rangle
}
is just the number of modes in the condensate band.
\item[ii)]Although all explicit projectors have been accounted for, the spatial integrals 
generate implicit projection
since we are working with a projected stochastic wavefunction $\alpha\equiv\PP\alpha$. 
\item[iii)]It is immediately apparent from \eref{Fadef} that the boundary terms vanish when 
either $[A,H_0]=0$, or 
\EQ{\QQ^*\left[(\delta
V(\mathbf{x},t)+u|\alpha(\mathbf{x})|^2)\;\alpha^*(\mathbf{x})\right]=0,}
which is automatically true if the modes at the energy cut-off are weakly occupied.
However, since the TWA has been used we have assumed that all modes in the
condensate band are significantly occupied. These two conditions can be reconciled if the occupation
of the modes at the cut-off is small relative to the other modes in the system but still large enough to render
the third order derivatives in the Fokker-Planck equation unimportant. 
\item[iv)]The advantage of neglecting the shape of the thermal cloud becomes clear when 
one considers the kind of terms that can show up in the spatially dependent calculation. 
A typical problematic term is proportional to $\QQ\bar{G}(\mathbf{x})\alpha(\mathbf{x})$. For a 
homogeneous cloud this term is zero, but when the spatial dependence becomes significant 
these terms, and additional ones that depend on $A\bar{G}(\mathbf{x})$ become important. 
\item[v)] Finally, we note the relationship between the number rate equation
\eref{dNdtsgpe} and the {\em simple growth
equation}~\cite{Gardiner1997,QKIII}
\EQ{\Label{simpgrowth}
\dot{N}_0=2W^+(N_0)\left\{(1-e^{(\mu_C(N_0)-\mu)/k_BT})N_0+1\right\}.
}
where $N_0$ is the condensate occupation number, $W^+(N_0)$ is the growth rate
into the condensate band, $\mu_C(N_0)$ is the condensate band chemical
potential, and $\mu$ is the non-condensate band chemical potential.  
To the degree of approximation we are using for
describing the non-condensate band we may linearize the exponential 
in \eref{simpgrowth} and make use of
\EQ{\Label{equivRates}
W^+(N_0)\equiv W^+=\gamma k_BT
}
to give the linearized simple growth equation
\EQ{\Label{simpgrowthlin}
\dot{N}_0=2W^+\left\{[\mu-\mu_C(N_0)]N_0/k_BT+1\right\},
}
while, from \eref{equivRates} and \eref{dNdtsgpe}, we find the {\em SGPE growth equation}
\EQ{\Label{simpgrowthSGPE}
\langle\dot{N}\rangle_{\scriptscriptstyle W}=2W^+\left\{\langle\mu-L_{GP}\rangle_{\scriptscriptstyle W}/k_BT+M\right\},
}
where $M\equiv\Tr\PP$ is the multiplicity of the condensate band. 
\par
We note two main differences between the two growth equations
\eref{simpgrowthlin} and \eref{simpgrowthSGPE}:
\begin{itemize}
\item The condensate chemical potential $\mu_C(N_0)$ of \eref{simpgrowthlin} is
replaced by a stochastic and spatial average of the $L_{GP}$ operator over the
condensate band. This is a significant generalization beyond the simple growth equation, and includes the
influence of random initial conditions without making the random phase approximation required to obtain
\eref{simpgrowth}. The great advantage of the SGPE approach is that it can describe both
fluctuations and coherences within the condensate band, and may also be a good description for 
quasi condensate growth where phase fluctuations dominate the condensate band evolution.
\item A more subtle difference arises in the spontaneous terms in \eref{simpgrowthlin} (given by the additional $+1$ in braces). 
This therm is a consequence of treating the condensate as a single mode
--that is, it is assumed to exist and be highly occupied \cite{QKIII},
and the rate equation describes those atoms in the condensate mode alone. 

The SGPE description is of the entire
condensate band, and the condensate fraction must be extracted from the full field. 
The spontaneous term in \eref{simpgrowthSGPE} (given by $+M$ in braces) arises from
the spontaneous growth into all of the single particle modes in the condensate band. 
\end{itemize}
\end{itemize}
\subsubsection{Bose-Einstein condensate in a harmonic trap}
There is an important simplification of the Ehrenfest relations for a potential that is 
parity conserving. The operator traces that arise directly from the noise in \eref{sgpedef}
are identically zero when the eigenstates conserve parity. This is easily seen for the
parabolic trap, the case of most experimental interest.
The driving terms in equations \eref{dxdtsgpe} and \eref{dpdtsgpe} are 
proportional to ${\rm Tr}(\PP {\bf x})$ and ${\rm Tr}(\PP {\bf p})$ respectively, consisting of 
sums of products of matrix elements of the form $\langle\phi_n|x|\phi_n\rangle$, and $\langle
\phi_n|p|\phi_n\rangle$. 
These terms vanish when $V_0(\mathbf{x})$ is parabolic because $x$ and $p$ couple harmonic oscillator
eigenstates $\phi_n(x)\to\phi_{n\pm1}(x)$. Similar considerations give $\Tr(\PP {\bf L})=0$, as
long as the thermal cloud is stationary in the laboratory reference frame\footnote{A
treatment of rotating frame condensation will be carried out in \cite{Bradley2004}}. If the
cut-off is chosen so that
the $\QQ$ projector terms are also small then we find the Ehrenfest relations for a trapped BEC
\EQ{
\fl\Label{dNdttrap}\frac{d \langle N \rangle_{\scriptscriptstyle W}}{dt}&=&2\gamma\langle
\mu-L_{GP}\rangle_{\scriptscriptstyle W}+2\gamma k_BT\Tr(\PP)\\
\fl\Label{dxdttrap}\frac{d\langle\mathbf{x}\rangle_{\scriptscriptstyle W}}{dt}&=&\frac{\langle\mathbf{p}\rangle_{\scriptscriptstyle W}}{m}+2\gamma{\rm Re}\langle \mathbf{x}(\mu-L_{GP})\rangle_{\scriptscriptstyle W}\\
\fl\Label{dpdttrap}\frac{d\langle\mathbf{p}\rangle_{\scriptscriptstyle W}}{dt}&=&-\langle\nabla V\rangle_{\scriptscriptstyle W}+2\gamma{\rm Re}\langle \mathbf{p}(\mu-L_{GP})\rangle_{\scriptscriptstyle W}\\
\fl\Label{dLdttrap}\frac{d\langle
\mathbf{L}\rangle_{\scriptscriptstyle W}}{dt}&=&-\frac{i}{\hbar}\langle\mathbf{L}V\rangle_{\scriptscriptstyle W}
+2\gamma{\rm Re}\langle \mathbf{L}(\mu-L_{GP})\rangle_{\scriptscriptstyle W}\\
\fl\frac{d\langle U\rangle_{\scriptscriptstyle W}}{dt}&=&\left\langle\frac{\partial V}{\partial
t}\right\rangle_{\scriptscriptstyle W}+2\gamma\langle
\overleftarrow{L}_{GP}(\mu-L_{GP})\rangle_{\scriptscriptstyle W}+2\gamma k_BT(\Tr[\PP(H_0+\delta V)]+u\langle\delta_C\rangle)\nonumber\\
\fl\Label{dEdttrap}
}
The fact that the finite temperature equations for $\langle {\bf x}\rangle, \langle {\bf p}\rangle$ and 
$\langle {\bf L}\rangle$ are the same as the equations 
found from \eref{dxdtsgpe}--\eref{dLdtsgpe} by setting $T=0$ (which amounts to neglecting the noise) has
interesting implications for condensate growth: the symmetry of
the trap plays a crucial role in isolating the condensate band kinematic degrees of
freedom from the direct influence of thermal noise. In a more general formulation
fluctuations would still enter through the inhomogeneity of the thermal cloud, but these results are a good description
in the high temperature regime.
\section{Projector terms for the harmonically trapped Bose gas} 
One of the main aims of the classical field method is to deal with finite temperature BEC's, but the method is constrained by the requirement that all modes in the condensate band are highly occupied. It is clear that if there is significant occupation near the cut-off the dynamics can be radically altered, but it is not sufficient to simply monitor the occupation numbers. It is preferable to find a strict {\em dynamical criterion} that ensures the validity of the simulations. Rather than tackle this general problem at finite temperature, in this work we will simply show that at zero temperature it is possible to use the Ehrenfest relations to construct a reliable estimator of the error arising from the cut-off for a particular process: the Kohn mode oscillation.
\par
To consider a simple example of the phase space boundary effects (given by $Q_A$ in
\eref{dNdtsgpe}--\eref{dEdtsgpe}) we consider a one dimensional model consisting of a harmonically trapped partially condensed Bose gas.
We also take $\gamma= T= 0$ so that $\langle \dot{N}\rangle=\langle \dot{E}\rangle= 0$, and\footnote[1]{We work in units of $x_0=(\hbar/m\omega)^{1/2}$, $t_0=\omega^{-1}$ and $k_0=1/x_0$ for length and time and wave-vector respectively.} 
\EQ{
\Label{pgpe}i\frac{\partial\psi}{\partial t}&=&\frac{1}{2}\left(-\frac{\partial^2}{\partial x^2}+x^2\right)\psi+\tilde{u}|\psi|^2\psi\\
\Label{dxdt1d}\frac{d\langle x\rangle}{dt}&=&\langle p\rangle+Q_{x}\\
\Label{dpdt1d}\frac{d\langle p\rangle}{dt}&=&-\langle x\rangle+Q_{p},
}
where $\tilde{u}$ is the dimensionless interaction strength. The details of how this relates to anisotropic trap geometry will not concern us here. We will merely consider modest effective nonlinearities of order $10< N\tilde{u}<1000$ to determine the validity of the criterion developed below.
\subsection{Harmonic oscillator basis}
We require some properties of the harmonic oscillator eigenstates. In terms of
the Hermite polynomials $H_n(x)$ the
eigenstates 
\EQ{
\phi_n(x)=\frac{H_n(x)}{\pi^{1/4}\sqrt{2^n n!}}e^{-x^2/2}
}
are coupled by the $x$ and $p$ operators to 
\EQ{\Label{H1}
x\phi_n(x)&=&\frac{1}{\sqrt{2}}\left(\sqrt{n}\phi_{n-1}(x)+\sqrt{n+1}\phi_{n+1}(x)\right)\\
\Label{H2} 
p\phi_{n}(x) &=& 
\frac{i}{\sqrt{2}}\left(\sqrt{n+1}\phi_{n+1}(x)-\sqrt{n}\phi_{n-1}(x)\right),
}
so that the projector generates the terms
\EQ{
\Label{Qxpsi}
\QQ \;x\psi(x)&=&\sqrt{\frac{N_c+1}{2}}\;c_{N_c}(t)\;\phi_{N_c+1}(x)\\
\Label{Qppsi}
\QQ \;p\psi(x)&=&i\sqrt{\frac{N_c+1}{2}}\;
c_{N_c}(t)\;\phi_{N_c+1}(x),
}
and the equations of motion are
\EQ{\Label{xho}
\frac{d\langle x\rangle}{dt}&=&\langle p
\rangle+{\rm Re}\;B(N_c+1)\\
\Label{pho}
\frac{d\langle p\rangle}{dt}&=&-\langle x\rangle+{\rm Im}\;B(N_c+1),
}
where
\EQ{
\Label{FN}B(n)&=&i\tilde{u}\sqrt{2n}c_{n-1}^*(t)\int
dx\;\phi_n^*(x)\;|\psi(x)|^2\psi(x).
}
As expected the correction is essentially a boundary effect caused by the
nonlinear term of the \GPE: this is the only term in the Hamiltonian that can cause transitions 
between states in the condensate and non condensate bands. Since $B(N_c+1)$ is proportional to $c_{N_c}^*(t)$ weak
occupation near the cut-off will naturally give Ehrenfest evolution. However, it is already apparent from the appearance 
of the nonlinear term that this condition alone is not sufficient to guarantee validity.
\begin{figure}[!ht]
   \begin{center}
               \mbox{
               \subfigure[Density time series]{\includegraphics[width=0.47\textwidth]{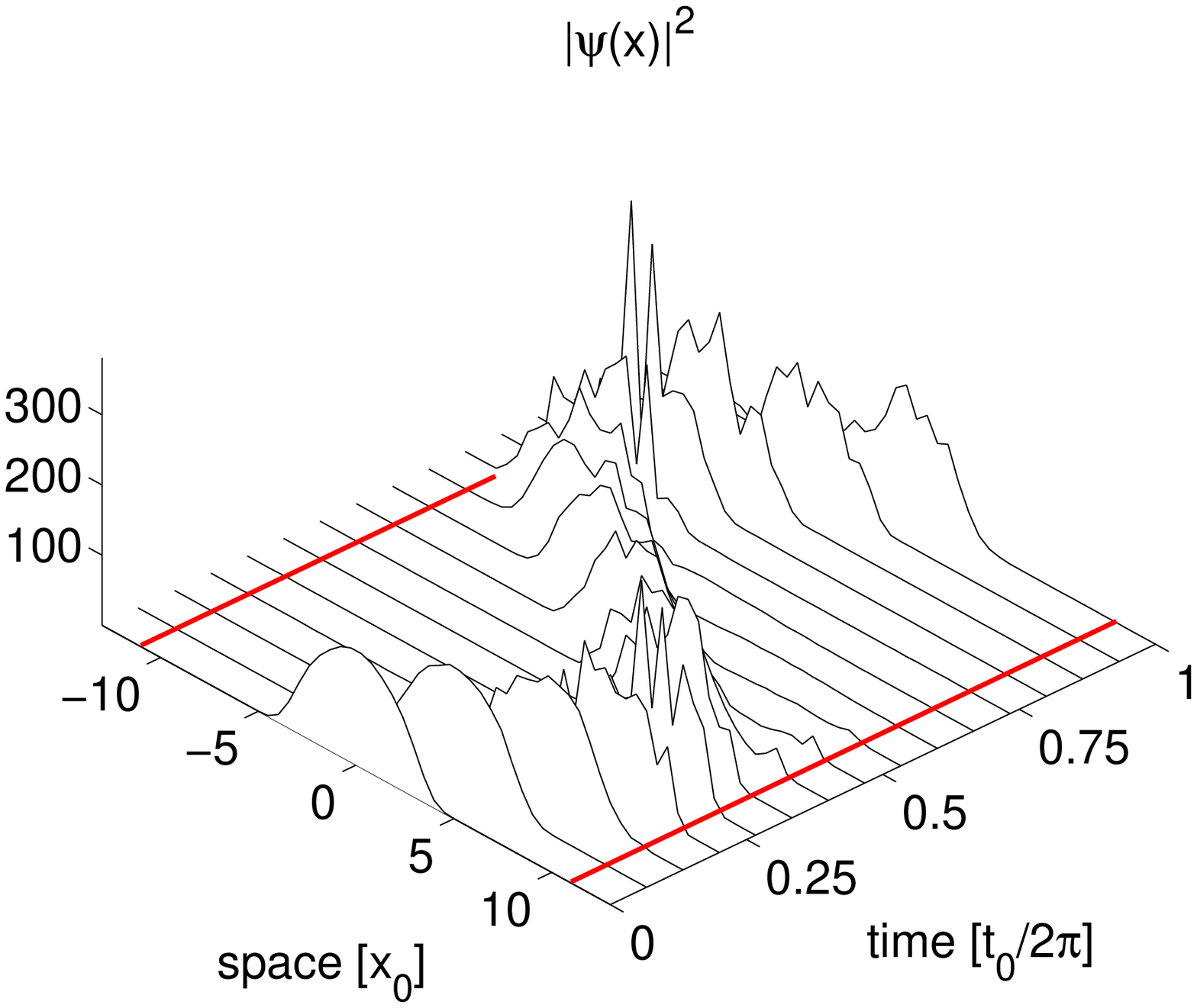}}
               \hspace{.1mm}
              \subfigure[PGPE ground state]{\includegraphics[width=0.47\textwidth]{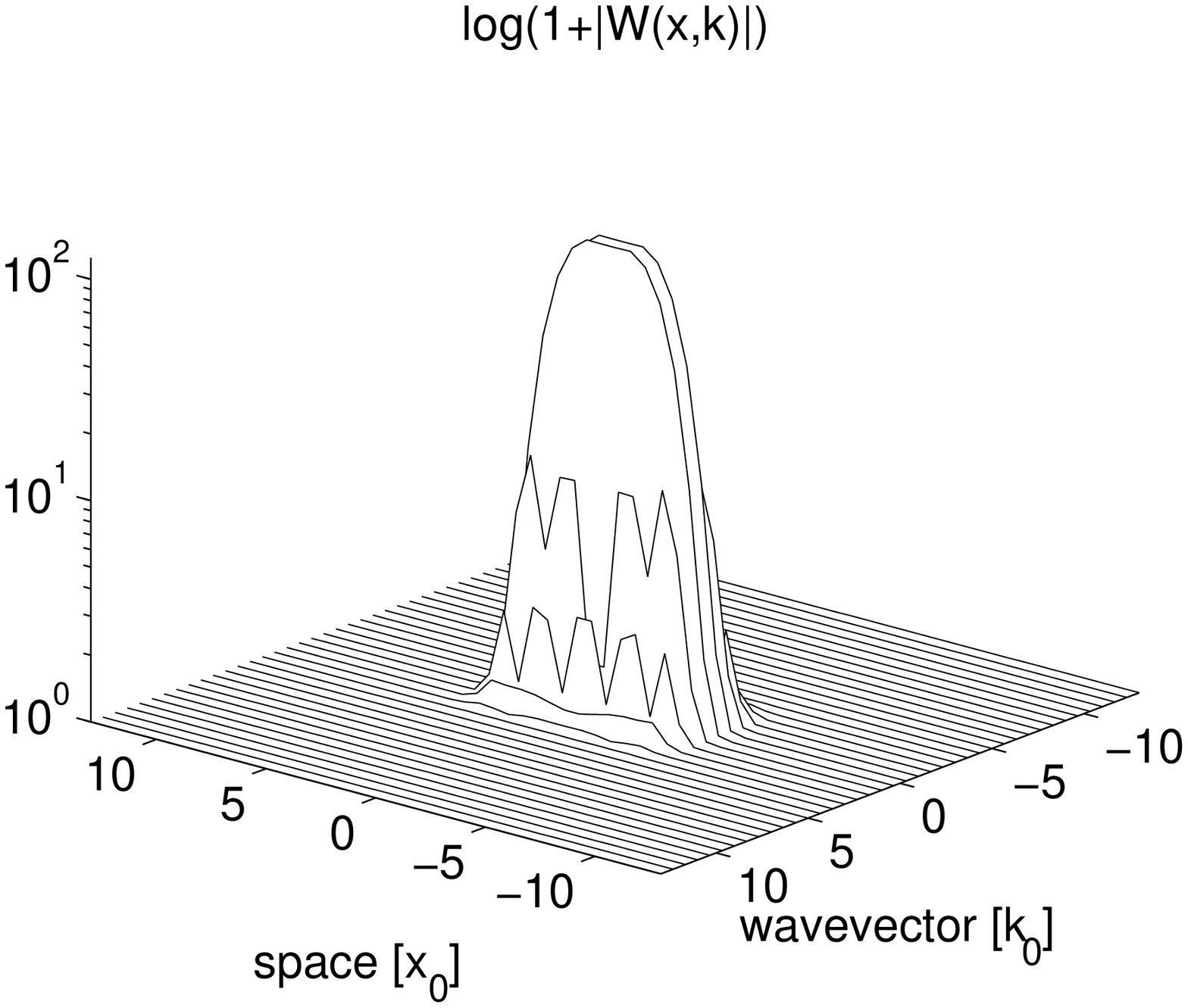}} 
               }
               \\
               \vspace{5mm}
               \mbox{
               \subfigure[Initial state shifted to $k_0=8$. ]{\includegraphics[width=0.47\textwidth]{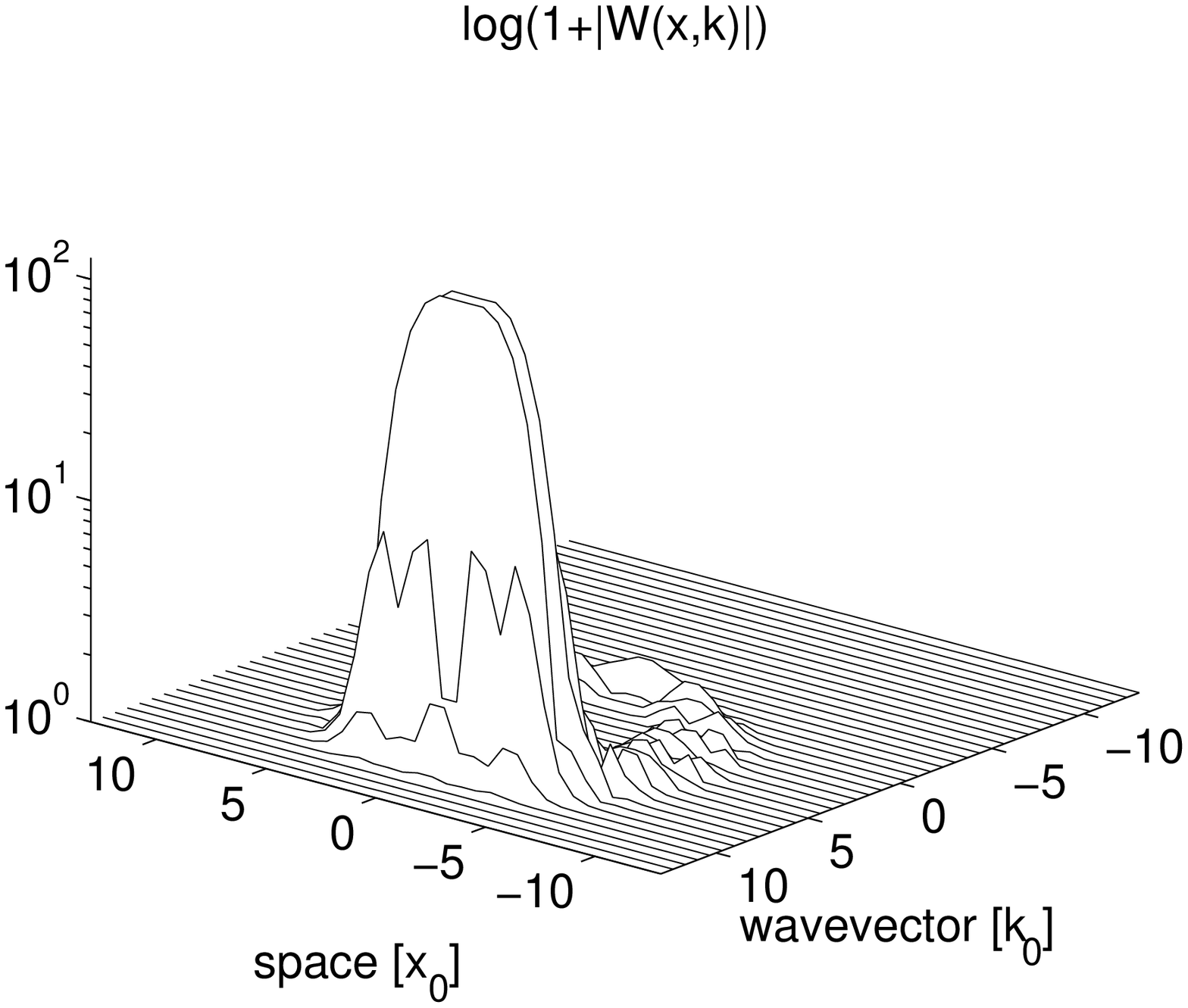}}
               \hspace{.1mm}
\subfigure[State after $1/4$ trap period]{\includegraphics[width=0.47\textwidth]{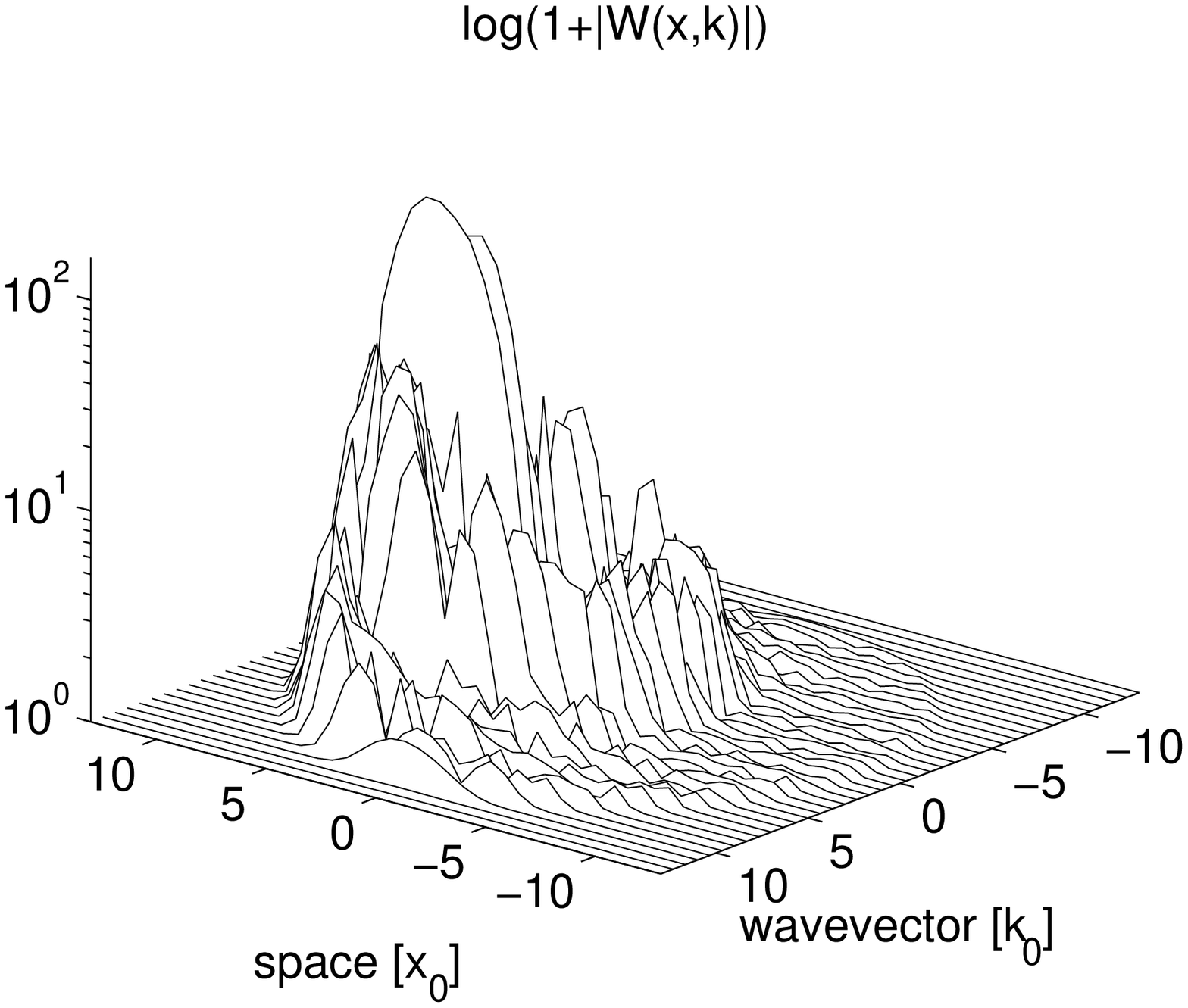}}
}     
               
               \caption{A simulation used to determine the threshold value of $E_z$ for a ground state wavefunction of the PGPE. (a) Distorted Kohn mode for motion near the semi-classical turning point of the cut-off mode (bold line). (b) Wigner function for the ground state solution of the fudge equation with nonlinear constant $\tilde{u}N=100$, cut-off $N_c=60$. (c) Projected initial state after a shift in momentum space to wave vector $k_0=8$. (d) At $1/4$ and $3/4$ of a trap period the projector has its strongest influence.}
               \label{typical_shape_oscillation}
               \end{center}
\end{figure}
We would like, therefore, to compute the relative error arising from the term $B(N_c+1)$ in the above equations. Since the Ehrenfest equations involve easily computed averages, one could simply monitor the ratios $|{\rm Re}\;B(N_c+1)/\langle p\rangle|$ and $|{\rm Im}\;B(N_c+1)/\langle x\rangle|$. Unfortunately, these are unsuitable as estimators because $\langle p\rangle$ and $\langle x\rangle$ regularly pass through zero. 
However, we can exploit the rotational symmetry of the cut-off in phase space by using the complex variable $\langle z\rangle\equiv\langle x\rangle+i\langle p\rangle$, for which the equation of motion is
\EQ{
\frac{d\langle z\rangle}{dt}=-i\langle z\rangle+B(N_c+1).
}
We can ensure that the cut-off is not altering the dynamics appreciably by requiring 
\EQ{
E_{z}\equiv\left|\frac{B(N_c+1)}{\langle z\rangle}\right|<10^{-4}.
}
\par
\begin{figure}[t]\centering
\includegraphics[width=0.7\textwidth]{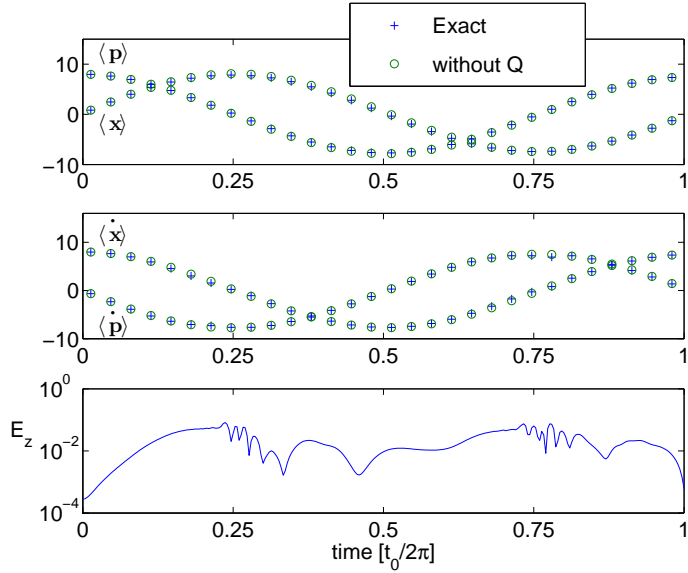}
               \caption{The position and momentum averages (top) and their time derivatives (middle) show almost complete immunity to the projector, although small deviations are detectable at at one quarter and three quarters of a trap period. The cut-off error estimator is significantly larger than $10^{-4}$ and has a well defined upper bound for a given initial energy.}
               \label{estimator}
\end{figure}

To demonstrate this we use the following test, which we believe to be rather stringent:
we take a ground state solution of the PGPE and give it a range of initial momenta in order to find the threshold where shape oscillations become apparent. For small kicks we observe Kohn mode oscillations and the projector has no effect on the dynamics. At a critical kick strength (determined by the cut-off energy), the edge of the condensate makes contact with the semi-classical turning point of the highest energy mode of the condensate band. The projector then comes into play, ensuring that the condensate wavefunction cannot make radial excursions in phase space that exceed the cut-off energy. This generates shape oscillations which become rather violent for large kick strength. A more revealing picture of the process is by transforming to phase space (which is detailed in the appendix), and the results for a high momentum simulation showing the distorted Kohn mode oscillations above threshold are presented in \fref{typical_shape_oscillation}. It is interesting to note that although the initial momentum kick has a negligible effect on the the shape of the condensate, the motion soon begins to fill the the available phase space --  a kind of spurious thermalisation is caused by the interplay between the process of nonlinear mode mixing and the cut-off.
The estimator we propose gives an unambiguous way of avoiding this predicament: it is clear from \fref{estimator} that the lowest order moments themselves are rather insensitive to the cut-off effects, nevertheless, $E_z$ is an accurate measure of the artifact.
\par
We can now map the threshold in two different ways: we examine the dependence of the threshold value of $E_z$ on the cut-off at fixed nonlinearity in \tref{ThresholdTablecut-off}, and find the variation with nonlinearity for a given cut-off in \tref{ThresholdTableNonlin}. 
\begin{table}
\begin{center}
\begin{tabular}{|c|c|c|}
	\hline
cut-off $N_c$ & $k_0$ & ${\rm max}[E_z]/10^{-4}$ \\ 
	\hline
30   & 1 & 20 \\
40   & 2.2 & 10 \\
50   & 2.5 & 2 \\
60  & 4 & 2 \\
100 & 7.3 & 1 \\
	\hline
\end{tabular}
\end{center}
\caption{
Threshold values for a range of cut-off energies. Other values are $\tilde{u}N=170$, $\mu = 20$.
\label{ThresholdTablecut-off}}
\end{table} 
\begin{table}
\begin{center}
\begin{tabular}{|c|c|c|}
	\hline
Nonlinearity $\tilde{u}N$ & $k_0$ & ${\rm max}[E_z]/10^{-4}$ \\ 
	\hline
59  & 9 & 2  \\
168  & 6.7 & 2  \\
476  & 4.8 & 8  \\
766 & 3.3 & 8 \\
1100 & 2.2& 10 \\
	\hline
\end{tabular}
\end{center}
\caption{
Threshold values for a range of nonlinear constants, with $N_c=100.$
\label{ThresholdTableNonlin}}
\end{table} 
The threshold value is independent of nonlinearity and cut-off energy over quite a wide range of either variable. It is important to note that, more generally, when any dynamical simulation is considered there will be a dominant set of moments that encapsulate the dynamics, and it is the effect of the projector on these quantities that must be considered. Higher order equations of the Ehrenfest type would then be used to extend the basic approach we have outlined here.
\section{Comparison of projector effects for the plane wave and harmonic oscillator bases for the harmonic trap}
The classical field method requires that the majority of modes used in the simulation of a Bose gas are highly occupied. 
When imposing this condition consistently near equilibrium in a harmonic trap it becomes clear that a strict energy cut-off should be used. Such a cut-off is best implemented in the basis of harmonic oscillator eigenstates, since, for a well chosen basis, the full interacting Hamiltonian for the finite temperature system is approximately diagonal at the cut-off energy. Since the plane wave basis is often used for classical field simulations of trapped systems it is important to evaluate the validity of such a procedure against the more accurate procedure based on an exact energy cut-off.
\par
We can use the expressions for the effect of the projector on equations of motion for averages to get an idea of 
the artifacts that arise when using the plane wave basis to represent a trapped system at finite temperature. 
\par
\subsection{Plane wave basis}
The system defined by $V_0(x,t)\equiv\delta V(x,t)\equiv 0$, with periodic boundary conditions
has been studied in great detail by Davis \etal\cite{Davis2002}. We treat the one dimensional case. 
\subsubsection{Plane wave basis without a trapping potential}
Writing
\EQ{\Label{pwdefn}
\psi=\sum_{k=-K}^{K}c_k\frac{e^{ikx}}{\sqrt{L}}
}
where $K$ is related to the spatial span according to
$K=2\pi N_c/L$. From this we have 
\EQ{
\QQ(x\psi)=\QQ\sum_{k=-K}^K c_k\frac{-i}{\sqrt{L}}\frac{\partial}{\partial k}e^{ikx},
}
which we can approximate by 
\EQ{
\QQ(x\psi)\approx\QQ\sum_{k=-K}^K c_k\frac{-i}{\sqrt{L}}\frac{e^{i(k+\Delta)x}-e^{i(k-\Delta)x}}{2\Delta},
}
when the momentum grid spacing $\Delta$ is small.
This leads to 
\EQ{\Label{Qx}\QQ(x\psi)\approx\frac{-i}{2\Delta\sqrt{L}}\left[c_K e^{i(K+\Delta)x}-c_{-K}e^{-i(K+\Delta)x}\right]
}
Since the momentum
operator commutes with the Hamiltonian and the potential is absent, the equations for the operator  
averages are $\dot{\langle N\rangle}=\dot{\langle E\rangle}=\dot{\langle p\rangle}=0$ and
\EQ{
\Label{boxx}\frac{d\langle x\rangle}{dt}&=&\frac{\langle p\rangle}{m}+Q_x(K)+S_\sigma(L)
}
where 
\EQ{\fl
Q_x(K)&=&\frac{u}{\hbar}2{\rm Im}\left\{\int dx\;|\psi|^2\psi^*\frac{-i}{2\Delta\sqrt{L}}\left[c_K(t) e^{i(K+\Delta)x}-c_{-K}(t)e^{-i(K+\Delta)x}\right]\right\},\\
\fl S_\sigma(L)&=&\frac{1}{2m}\left[(x\psi)(p\psi^*)-\psi^*(px\psi)\right]^{L/2}_{-L/2}.
}
The boundary term $S_\sigma$ arises from the finite span of the periodic basis, and 
will be important in situations where $\psi$ is non-zero at the box boundaries. This 
term does not arise in the use of the spectral basis, since the basis elements are 
defined over all space. 

\subsubsection{Modeling a quadratic potential with a plane wave basis}
We have seen that there is a boundary term for the $\dot{\langle x\rangle}$ equation 
when the plane wave basis is used for $V(x,t)\equiv0$. When the plane wave basis is 
used to model a system with a harmonic potential, the results follow from  
\eref{Fadef} by putting $\delta V(x)=m\omega^2x^2/2$, so that the entire potential becomes 
a variation with respect to $H_0$. 
The continuity equation becomes
\EQ{\Label{contplane}
\frac{\partial n(\mathbf{x})}{\partial t}+\nabla \cdot\mathbf{j}(\mathbf{x})=\frac{2}{\hbar}{\rm Im}\left(\QQ^*\left[(m\omega^2x^2/2+u|\psi|^2)\psi^*\right]\psi\right),
}
and the Ehrenfest relations become
\EQ{
\Label{shox}\frac{d\langle x\rangle}{dt}&=&\frac{\langle p\rangle}{m}+Q_x(K)\\
\Label{shop}\frac{d\langle p\rangle}{dt}&=&-m\omega^2\langle x\rangle\\
}
with the boundary term given by
\EQ{
\fl\Label{Qxsho}Q_x(K)&=&\frac{1}{\hbar}\;2{\rm Im}\Bigg(\int
dx\;\left(m\omega^2x^2/2+u|\psi(x)|^2\right)\psi^*(x)\\
\fl&&\times\frac{-i}{2\Delta\sqrt{L}}\Big[c_K(t) e^{i(K+\Delta)x}-c_{-K}(t)e^{-i(K+\Delta)x}\Big]\Bigg).
}
We have neglected $S_\sigma$ because the trapping potential will keep the wavefuction negligible at the grid
edge for a well chosen basis.
Comparing the plane wave basis Equations \eref{shox}--\eref{Qxsho} 
with the spectral basis results \eref{xho}--\eref{FN} 
(for $\delta V\equiv0$), we see that there are two notable differences: i) The boundary 
corrections occur in both the $\dot{\langle x\rangle}$ and $\dot{\langle p\rangle}$ 
equations for the spectral basis, and only in the $\dot{\langle x\rangle}$ equation 
in the plane wave basis; and ii) there is a contribution from the potential in the boundary 
correction for $\dot{\langle x\rangle}$ for the plane wave basis. This is 
particularly significant since this term can potentially assume large values when 
$c_K(t)\neq 0$, {\em even in the linear regime}. 

\subsection{The optimal plane wave representation}
There is a certain degree of freedom in choosing a plane wave basis for representing a harmonically 
trapped system. Here we show how to obtain the optimal plane wave basis that best captures the 
lowest harmonic oscillator states. We would expect this to be the best plane wave representation for 
modeling harmonically trapped systems. 

We consider a basis of $N_{c}$ plane wave states taken to extend over the spatial box of size 
$x\in[-L/2,L/2]$, as defined in \eref{pwdefn}. For fixed $N_{c}$, the only free parameter 
in constructing the plane wave basis is $L$. Making $L$ large is done at the expense of decreasing 
the momentum width that can be represented on the grid, while conversely decreasing $L$ limits the 
spatial extend of the system, but increases the momentum range. 
 
Here we give a simple argument for an \emph{optimal} choice of $L$ at fixed $N_{c}$. 
In harmonic oscillator units the single particle Hamiltonian for the 
harmonic oscillator takes the  form
\begin{equation}
\bar{H} =\frac{1}{2}\bar{k}^2+\frac{1}{2}\bar{x}^2,
\end{equation}
where bars are used to indicate dimensionless quantities (e.g. see Sec. VB1 of \cite{Cohen77}). 
In these units the Hamiltonian and its eigenstates take the same form in coordinate and wave 
vector space. So the best grid choice will be when our numerical grids for (dimensionless) 
position and wave vector space are identical, i.e. when $\bar{L}=\bar{K}$. Returning to 
dimensioned units this optimal choice is
\begin{equation}
L_{\rm{opt}}=\sqrt\frac{2\pi \hbar N_{c}}{m\omega}.
\end{equation}
From this expression we directly obtain that the largest momentum the can be represented on the 
grid is $K_{\rm{opt}}=\pi N_{c}/L_{\rm{opt}}$ (i.e. the limits of the sum in \eref{pwdefn}).

\begin{figure}[!htb]{\centering 
\includegraphics[width=3.1in]{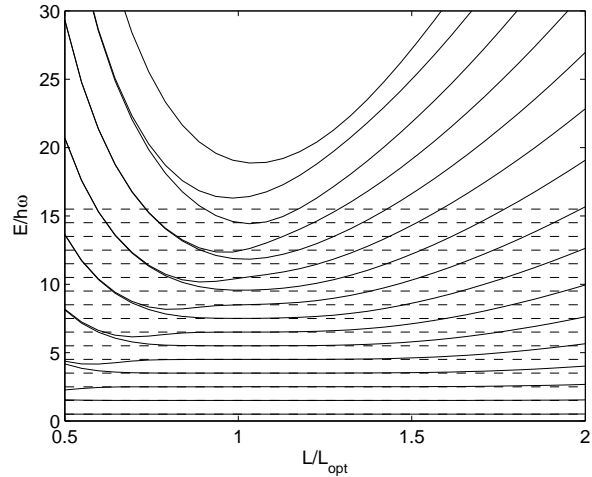}\par}
\caption{The numerical spectrum of the harmonic oscillator Hamiltonian. The solid lines are 
the plane wave results found by diagonalizing the harmonic oscillator Hamiltonian on a grid 
of $N_{c}=16$ points of width $L$. The dashed lines indicate the $16$ lowest energies of the 
exact eigenspectrum, which corresponds to diagonalising the Hamiltonian in the spectral basis 
for $N_c=16$.  
\label{Spectrum}}
\end{figure}

To quantify the sensitivity to non-optimal choices of $L$, in \fref{Spectrum} we show the 
spectrum of energies found by diagonalizing the harmonic oscillator Hamiltonian in the plane 
wave basis for a range of $L$ values. These results show that $L_{\rm{opt}}$ is clearly the 
best choice, however even for $L=L_{\rm{opt}}$ only about half the eigenstates are accurately 
obtained.   

\subsection{Comparison of plane wave and harmonic oscillator phase space}
\begin{figure}[!htb]{\centering 
\includegraphics[width=2.1in]{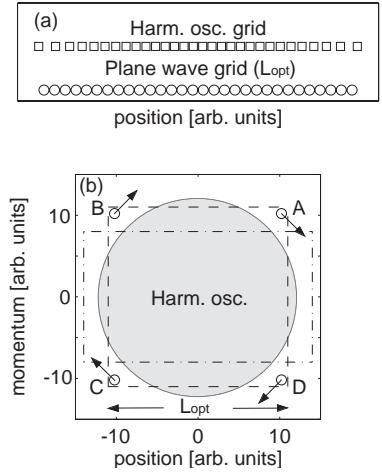}\par}
\caption{Phase space of the harmonic oscillator and plane wave representations. (a) The \emph{quadrature} grids for the 
optimal plane wave basis (circles) and the harmonic oscillator basis (squares). (b) The approximate phase space captured by 
the optimal plane wave (dashed boundary), a non-optimal plane wave (dash-dot boundary), and harmonic oscillator (solid boundary) bases.  
The points $A$-$D$ indicate the evolution of anomalous trajectories (see text).
\label{PhaseSpace}}
\end{figure}

The harmonic oscillator and plane wave bases differ somewhat in the regions of phase space 
they represent. This difference is reflected in the position quadrature grids associated with 
each basis, shown in \fref{PhaseSpace}(a). It is apparent from this figure, that the 
spacing between quadrature points of the harmonic oscillator basis varies from being dense in 
the central region to sparse at the edges. This enables the basis to capture large momentum 
states (i.e. fast spatial variations) at small displacements for the trap center, and smaller 
momentum states at large displacements. This suggests that for fixed $N_c$, the harmonic oscillator 
basis captures a circular region of phase space, as shown schematically in \fref{PhaseSpace}(b). 
In contrast, the plane wave grid is equally spaced over the entire region 
$[-L/2,L/2]$ (see \fref{PhaseSpace}(a)). This means the plane wave representation is equally well 
able to represent high momentum states at all displacements from the trap centre, suggesting 
that this basis captures a rectangular region of phase space (see \fref{PhaseSpace}(b)). 
\par
Classically the motion of a harmonic oscillator corresponds to perfect circular trajectories 
in phase space, and we expect that a circular phase space projector forms the ideal energy cut-off. 
We note that the optimal plane wave representation corresponds to choosing $L$ so that the 
maximum kinetic and potential energies associated with the edge values of the position and 
wave vector grids respectively are equal, i.e.  
$\frac{1}{2}m\omega^2(L_{\rm{opt}}/2)^2=\hbar^2 (K_{\rm{opt}}/2)^2/2m$. For this case the 
phase space region bounded by the plane wave representations most closely matches the harmonic 
oscillator space (see the dashed line in \fref{PhaseSpace}(b)). In comparison, a non-optimal 
plane wave basis has an energy projector that restricts the kinetic energy and potential energy inconsistently, 
giving rise to a rectangular phase space boundary (e.g. see the dash-dot line in \fref{PhaseSpace}(b)).
\par
The high energy modes in the plane wave phase space exhibit anomalous dynamics arising from aliasing 
the region of phase space that is inconsistently represented. We illustrate this by examining the dynamics 
of a phase space point $A$ in \fref{PhaseSpace}(b). This point evolves along the trajectory 
indicated by the arrow until it reaches the right position boundary (dashed line). It is then aliased 
to the left position boundary and continues to evolve through point $B$, before reaching the upper 
momentum boundary. It will then pass through point $C$, $D$ and then return to $A$. The overall result 
is that states lying near the corners of the phase space region undergo a counter-clockwise evolution 
through phase space, in contrast to the normal clockwise evolution through phase space. In application 
to the SGPE formalism, we would expect that system disturbances with momentum and position characteristics 
lying in these corner regions will evolve in this anomalous manner. 

\section{Comparison of plane wave and spectral representations for a thermalized 1D gas \label{SecTherm}}
In this section we compare the effect of basis on simulations of a harmonically trapped 1D gas 
using the PGPE equation (\ref{PGPE}). Our method follows the approach used by Davis {\em et al.} 
in reference \cite{Davis2002}: For each of the bases under consideration we evolve a randomized 
initial state of definite energy according to the PGPE. This evolution is expected to be ergodic, 
and by appropriately time averaging pure state expectations we are able to obtain ensemble averages. 
To compare the different bases we examine the equilibrium position and momentum density distributions, 
and the condensate fractions. 

\begin{figure}[!htb]{\centering 
\includegraphics[width=4in]{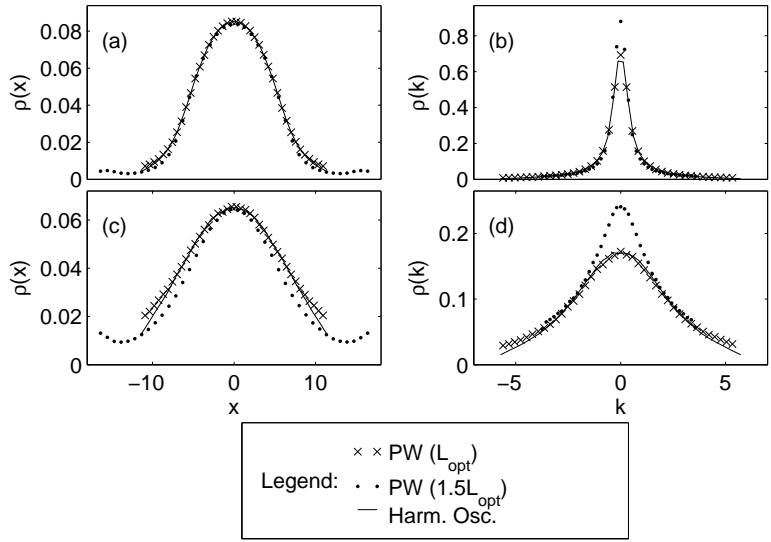}\par}
\caption{Results for the equilibrium position and momentum density profiles of a 1D thermal Bose gas. 
{\bf Low Energy Case:} Simulations for a total energy of $E=14\hbar\omega$ (a) equilibrium position 
density, (b) equilibrium momentum density. {\bf High Energy Case:} Simulations for a total energy of 
$E=21\hbar\omega$ (c) equilibrium position density, (d) equilibrium momentum density.
Results are calculated by time-averaging classical field calculations carried out in the different 
bases under consideration: plane wave for  $L=L_{\rm{opt}}$, $L=1.5L_{\rm{opt}}$, and the harmonic 
oscillator basis. Other simulation parameters: $u=200x_0/\omega$, $N_{c}=40$.
\label{ThermalizeFig}}
\end{figure}

In detail, the simulations we have conducted are for a dimensionless interaction strength of 
$u=200x_0/\omega$ for bases with 40 modes (i.e. $N_{c}=40$). 
In Fig. \ref{ThermalizeFig} we present results for the density distributions found from evolving randomized initial states with a total energy of  $E=14\hbar\omega$  
(Figs. \ref{ThermalizeFig}(a) and (b)), and for $E=21\hbar\omega$ (Figs. 
\ref{ThermalizeFig}(c) and (d)). These two choices of energy correspond to a strongly condensed system (with a 
large condensate fraction) and a system close to the transition respectively 
(we will discuss condensate fraction later in this section). The three bases we compare are 
the spectral basis, and plane waves bases with $L=L_{\rm{opt}}$, $L=1.5L_{\rm{opt}}$.  

The results in \fref{ThermalizeFig} confirm that the the optimal plane wave basis is in better 
agreement with the spectral bases than the non-optimal basis.  
A particular weakness of the plane wave representations is that density distributions in the wings tend to 
decrease more slowly than the spectral basis. In the $1.5L_{\rm{opt}}$ case the density distribution even begins to increase near the boundary, as is apparent in 
figures \ref{ThermalizeFig}(a) and (c). This behavior is most likely due to the inconsistent manner 
that the plane wave basis represents the highest energy states as discussed in the previous 
subsection. This will have serious implications for schemes that use the behaviour of the distribution 
wings to calculated temperature, and is likely to have effected the temperature calculations made in \cite{Goral2002}.

Finally we examine how the condensate fraction is influenced by the choice of basis. We determine 
the condensate fraction using the Penrose-Onsager criterion \cite{Penrose1956,Goral2002,Blakie2004A}. 
To do this we calculate the 1 body density matrix by time-averaging the classical field, i.e. 
$\rho_{\rm{1B}}(x,x^\prime)=\langle \psi^*(x)\psi(x^\prime)\rangle_{\rm{time ave.}}$. The condensate 
fraction is determined as the largest eigenvalue of the 1 body density matrix. The results are 
presented in \tref{CondFracTable} for the cases considered in \fref{ThermalizeFig}, augmented by 
the results from a wider range of plane wave bases. 
These results show conclusively that non-optimal plane wave bases can have a dramatic influence 
on the physical properties of the system simulated.
\begin{table}
\begin{center}
\begin{tabular}{|c|c|c|}
	\hline
Basis &  Cond. Frac. $E=14\hbar\omega$    &  Cond. Frac.  $E=21\hbar\omega$     \\ 
	\hline
H. Osc.   &  0.370 & 0.072   \\
PW ($0.5L_{\rm{opt}}$)   &  0.815 & 0.238 \\
PW ($0.8L_{\rm{opt}}$)    &0.392& 0.097  \\
PW ($1.0L_{\rm{opt}}$)   & 0.370 &  0.078 \\
PW ($1.2L_{\rm{opt}}$)    &  0.360 & 0.078  \\ 
PW ($1.5L_{\rm{opt}}$)   & 0.444 &  0.116  \\ 
	\hline
\end{tabular}
\end{center}
\caption{
Condensate fraction for a 1D thermal Bose gas obtained using various numerical bases.
\label{CondFracTable}}
\end{table} 

\section{Conclusions}
We have shown how to derive exact Ehrenfest relations for the PGPE, an equation of motion
which has become an important tool in the study of finite temperature Bose-Einstein 
condensates. The Ehrenfest relations show the interesting feature that for a harmonically 
trapped BEC in contact with a high temperature thermal cloud the operator averages for the 
kinematic degrees of freedom are immune to thermal noise {\em on the average}. 
\par

For simulations of 
the PGPE and the SGPE in the truncated Wigner or classical field approximation it is required 
that all modes in the condensate band are significantly occupied 
\cite{Sinatra2002,SGPEII,Davis2001a}, however it is the {\em relative occupation} 
at the phase space boundary that determines the influence of the projector. Thus the cut-off can be chosen so that the projector 
does not generate spurious dynamics, even though the modes near the cut-off may have moderate occupation. However, this can be a delicate balance, and we have shown how to find easily computable {\em dynamical} tests of the PGPE that are not reliant on simply monitoring the mode occupation numbers.
\par
The projector in the SGPE generates boundary terms that arise because the Gross-Pitaevskii
time evolution can evolve the wavefunction outside the condensate band. This kind of 
evolution can arise through either the nonlinear term or from an additional potential 
which is not part of the single particle Hamiltonian used to generate the representation basis. 
It therefore becomes important to choose the right basis, and we 
have shown that using the plane wave basis for a harmonically trapped BEC in thermal equilibrium can significantly alter the equilibrium condensate
fraction unless the basis is chosen to optimally reproduce the single particle spectrum of the harmonic trap.
\ack 
AB thanks Murray Olsen, Matthew Davis, Piyush Jain and Adam Norrie for 
many usefull comments.
\begin{appendix}
\section{Transforming to phase space}\label{app:A}
In order to examine the phase space behaviour of the condensate band
wavefunction we require the connection between the mode representation
$\psi(x)=\sum_n c_n\phi_n(x)$ and the Wigner function
\EQ{\Label{Wdef}
W(x,k)=\frac{1}{2\pi}\int dy\;e^{iky}\psi^*(x+y/2)\psi(x-y/2)
}
for the harmonic oscillator basis. 
\par
We insert the mode decomposition into \eref{Wdef} to get
\EQ{
W(x,k)=\sum_{n,m}^{-}c_n^*c_m\;W_{nm}(x,k),
}
where the modes $\phi_n(x)$ are the orthonormal eigenstates of the harmonic trap and
\EQ{
W_{nm}(x,k)\equiv\frac{1}{2\pi}\int dy\;e^{iky}\phi_n(x+y/2)\phi_{m}(x-y/2).
}
For brevity we will use the notation
\EQ{\Label{Wnqdef}
W_n^q(x,k)= W_{n,n+q}(x,k).
}
\par
We make use of the contour representation of the Hermite polynomials \cite{A&W} to write 
the modes
as
\EQ{
\phi_n(x)=\frac{e^{-x^2/2}}{\sqrt{2^nn!\sqrt{\pi}}}\oint \frac{dt\;e^{-t^2+2tx}}{t^{n+1}}
}
and \eref{Wnqdef} becomes
\EQ{
W_n^q(x,k)=&\frac{e^{-x^2}}{2\pi(2\pi i)^2}\sqrt{\frac{n!(n+q)!}{2^{2n+q}\pi}}\oint\frac{ds\;e^{-s^2+2sx}}{s^{n+q+1}}\nonumber\\
&\times\oint\frac{dt\;e^{-t^2+2tx}}{t^{n+1}}\int dy\;e^{-y^2/4+y(ik+t-s)}
}
where the contours enclose the origin.
Carrying out the $y$ and $t$ integrals gives
\EQ{\fl
W_n^q(x,k)=\frac{e^{-x^2-k^2}}{\pi}\sqrt{\frac{n!2^q}{(n+q)!}}\frac{1}{2\pi i}\oint\frac{ds\;e^{2s(x+ik)}}{s^{n+1}}(x-ik-s)^{n+q},
}
which, after the change of variables $s=z(x-ik)/(z-1)$, becomes
\EQ{\fl
W_n^q(x,k)=\frac{(-1)^ne^{-x^2-k^2}}{\pi}\sqrt{\frac{n!2^q}{(n+q)!}}(x-ik)^q\frac{1}{2\pi
i}\oint\frac{dz}{z^{n+1}}\frac{e^{-2(x^2+k^2)z/(1-z)}}{(1-z)^{q+1}}.
}
We can now use the generating function for the associated Laguerre polynomials \cite{A&W}
\EQ{
\frac{e^{-xz/(1-z)}}{(1-z)^{q+1}}=\sum_{n=0}^\infty z^nL_n^q(x)
}
to find
\EQ{
W_n^q(x,k)=\frac{(-1)^n}{\pi}\sqrt{\frac{n!2^q}{(n+q)!}}e^{-x^2-k^2}(x-ik)^qL_n^q(2(x^2+k^2)).
}
When $q=0$ we recover the Wigner function for a number state which is a
well known result in quantum optics \cite{QN}.
\par 
Using the symmetry $W_{n+q,n}(x,k)=W_n^q(x,k)^*$, the transformation to phase space is
\EQ{\fl
W(x,k)=2{\rm Re}\left\{\sum_{q=0}^N\sum_{n=0}^{N-q}c_n^*c_{n+q}W_n^q(x,k)\right\}-\sum_{n=0}^N|c_n|^2W_n^0(x,k).
}
\end{appendix}
\section*{References}
\bibliographystyle{prsty}
\bibliography{References}

\begin{thebibliography}{10}
\bibitem{SlowSpec}
Although the basic algorithms are not as adaptable as Fourier methods, for a
  given trap geometry a spectral algorithm can be quite fast because one does
  not need to use a lot of single particle states compared to the large set of
  plane waves needed to describe the same system.
\end{thebibliography}
\end{document}